\documentclass[prb,aps,twocolumn,showpacs,floatfix,superscriptaddress]{revtex4}
\newcommand{\secc}[1]{\section{#1}}

\usepackage{graphicx,times}
 \newcommand{\MEMO}[1]{}
\newcommand{\KEEP}[1]{{#1}}
\newcommand{\OMIT}[1]{}
\newcommand{\OMITX}[1]{}

\newcommand{\SAVE}[1]{}

\begin{document}

\title{Empirical oscillating potentials for alloys from ab-initio fits and
the prediction of quasicrystal-related structures in the Al--Cu--Sc system}

\author{Marek Mihalkovi\v{c}}
\affiliation{Laboratory of Atomic and Solid State Physics, Cornell University,
Ithaca, NY, 14853-2501}
\affiliation{
Permanent address,
Institute of Physics, Slovak Academy of Sciences, 84228 Bratislava, Slovakia.}
\author{C. L. Henley}
\affiliation{Laboratory of Atomic and Solid State Physics, Cornell University,
Ithaca, NY, 14853-2501}

\begin{abstract}
By fitting to a database of ab-initio forces and energies, 
we can extract pair potentials for alloys, with  a simple six-parameter
analytic form including Friedel oscillations, which give
a remarkably faithful account of many complex intermetallic
compounds.  
  Furthermore, such 
potentials are combined with a method of discovering
complex zero-temperature structures with hundreds of atoms
per cell, 
given only the composition and the constraint of known lattice
parameters, using molecular--dynamics quenches.
We apply this approach to structure prediction 
in the Al--Cu--Sc quasicrystal-related system.
\end{abstract}

\pacs{02.70.Ns,61.50.Lt,63.20.Dj,64.70.Kb,61.44.Br}

\maketitle

\MEMO{TODO-0: Review it overall, are any sections obsolete
in a subtle way (emphasis related to old versions)?}

Various problems in materials modeling can only be 
addressed by ``empirical'' interatomic potentials~\cite{FN-empirical}.
Say we wish to evaluate a physical property
(e.g. total energy) of some material with a complex crystal structure.
We cannot directly insert the results of crystallographic
refinements, as they (almost always) include sites with 
mixed or fractional occupancies.
To obtain valid results, we must assign those occupancies
plausibly based on computed energy differences.
A single relaxation with fast ab-initio codes such as VASP~\cite{vasp}
may be tractable in a cell of $10^3$ atoms;
however, if (say) 5\% of those atoms are uncertain, 
this must be repeated many times to find the one optimum
state out of $2^{50}$ possibilities.

But with empirical, approximate potentials that can be evaluated 
in negligible time, this optimization is tractable, and can be
followed up by ab-initio relaxation to obtain the most
accurate positions and total energies.
When combined with effective search algorithms,
such as genetic algorithms or the ``cell--constrained
melt--quench'' method (presented in Sec.~\ref{sec:constrained})
this is a powerful tool for ab-initio structure discovery.
Some other questions that call for empirical potentials are 
phonon spectra (or other dynamics),  and thermodynamic simulations
of phase transformations in complex alloys.

\SAVE{Not mentioned: Material design/modelling, bridging 
microscopic-to-macroscopic scales. Nanocrystals.
As for dislocations, maybe not valid.}

\MEMO{can we just say ``embedded-atom method'' (EAM)~\cite{EAM,MEAM}?}
A popular framework of empirical potentials
is the ``embedded-atom method'' (EAM)~\cite{EAM,MEAM},
in which the full Hamiltonian contains the usual pair term $V_{ij}(r_i-r_j)$, 
but also an implicitly many-atom term $\sum_i U(\rho(r_i))$, where 
$\rho(r_i)$ is a sum of contributions at atom $i$ from nearby atoms.
Accurate EAM potentials are straightforward to extract
for pure elements,
but demand patience and skill to obtain even for binary systems;
there is no 
{\em systematic} recipe for multicomponent systems.

Here we present an alternative approach
fitting only pair interactions but incorporating Friedel oscillations.
These ``empirical oscillating pair potentials'' (EOPP)
have the form~\cite{suppl}
   \begin{equation}
   \label{eq:oscil6}
    V(r) = \frac{C_1}{r^{\eta_1}} + \frac{C_2} {r^{\eta_2}} \cos(k_* r + \phi_*)
   \end{equation}
\SAVE{Relation to Marek's internal form (with $\{ a_i \}$):
$C_1=a_0^{a_1}$; $\eta_1=a_1$; 
$C_2=a_2$; $\eta_2=a_5$;
$k_*=a_3$; $\phi_*=a_4$.}
All six parameters, including \OMIT{the spatial frequency} $k_*$,
are taken as independent in the fit for each pair of elements. 
Eq.~(\ref{eq:oscil6}) was inspired by effective potentials
(e.g. Refs.~\onlinecite{hafner} and \onlinecite{widom-GPT}) 
used in previous work
on structurally complex metals, e.g.
quasicrystals~\cite{Al-TM,krajci-AlZnMg}.
In such systems, energy differences between competing structures
are often controlled by second- and third-neighbor wells due to
Friedel oscillations, 
\KEEP{which are a consequence (mathematically)
of Fourier transforming the Fermi surface, or (physically) are
equivalent to the Hume-Rothery stabilization by enhancing
the strength of structure factors that hybridize states
across the Fermi surface.}%
~\cite{pettifor}
In Pettifor's framework (Ref.~\onlinecite{pettifor}, Sec. 6.6),
the short-range repulsion is captured by the first term of
(\ref{eq:oscil6}); the medium-range potential (first-neighbor well) 
as well as the long-range oscillatory tail are captured by
the second term, their relative weights being adjusted by
the $\eta_1$ and $\eta_2$ parameters.  
Since the second term has this double duty of representing
both the nearest-neighbor and long-distance behaviors, the 
fitted $1/r^{\eta_2}$ decay generally does not agree with
analytic asymptotic result of $1/r^3$; also, $k_*$ need not
match the Fermi wavevector, and these parameters take
different values for the six kinds of pair potentials.

In the rest of this Letter, we describe how our potentials
are fitted (typically for a particular composition range)
and then demonstrate their capabilities 
through case studies in the alloy system  Al--Cu--Sc,
which has local order similar to the binary quasicrystal
$i$(CaCd) or to Zn-rich Mg--Zn alloys.
Along with this, we also describe a method of ``constrained
cell quenching'' that accesses low-energy structures
in surprisingly large cells.
Finally, we will  summarize other systems where EOPP have
been applied, and discuss their limitations.

\MEMO{TODO. Need to emphasize better here: this is a general recipe!}

\secc{Database and potential fitting}
The parameters in Eq.~(\ref{eq:oscil6}) are fitted to an ab-initio dataset
(using the VASP code~\cite{vasp})
combining both relaxed $T=0$ structures and molecular dynamics 
(MD) simulations at high $T$.
Four criteria in choosing structures for the database are
(1) to bracket the composition range of interest;
(2) to mix simple and complex structures;
(3) to ensure adequately many contacts of
every kind (in particular, nearest neighbors of
the least abundant species);
(4) to have similar atom densities~\cite{fn-densityrange}.
In ab-initio MD simulations of the simpler structures, a
supercell is always used with dimensions comparable to the
fitting potential cutoff radius, which  (for the fit procedure)
is  always 12\AA~\SAVE{and abrupt}.
\SAVE{(However, we used 9 \AA~in the actual melt-quench simulations.)}

\SAVE{In our Sc--Zn study,the database used  
binaries UHg$_2$, MgZn$_2$,  ScZn$_6$,  and Mg$_4$Zn$_6$.cI160,
(with U$\to$Sc,Hg$\to$Zn, Mg$\to$Sc).
For the MD (force) data, we used $T_{\rm MD} \ge 800$K.
The database for ZnSc included 4035 force points and 74
relaxed energies; 
the r.m.s.(forces) was 0.16 eV/\AA~ and
r.m.s.(energies) was 7.9 meV/atom.
Energies are always expressed as a difference from the tie-line or tie-plane
defined by the pure elements
[Our 2007 version of this paper said $T_{\rm MD}\approx$ 300 K 
for Zn--Sc phases; it also mentioned ``$T$(Zn$_{96}$Sc$_{64}$)'' 
in the ``[$T$(AlMgZn)]'' structure, as part of the database.]
}

\begin{figure}
\includegraphics[width=3.4in]{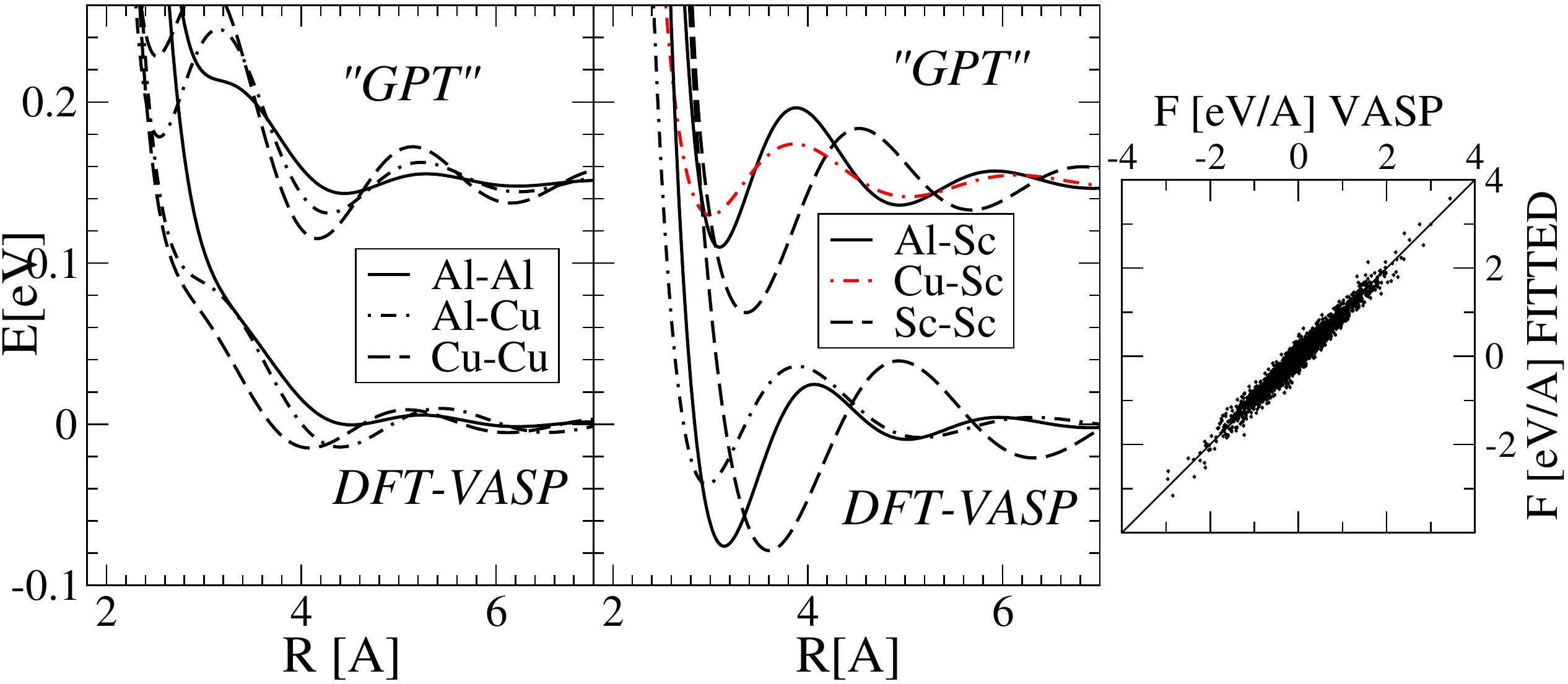}
\caption{
(a). {\it Top}: EOPP form (Eq. \ref{eq:oscil6}) fitted to ``GPT'' potentials  
(Ref.~\onlinecite{moriarty-GPT}); the fitted/GPT curves lie exactly on
the top of each other. 
The Cu--Sc potential is not available from Ref.~\onlinecite{widom-GPT};
the figure shows a fit to ab--initio data with the other five pairs
constrained to GPT form. {\it Bottom}: the EOPP form fitted to VASP force and
energy data. The curves fitted to ``GPT'' are shifted by $0.15$ 
eV/atom for clarity. 
\OMIT{ MMNEW: below is previous caption text.
(a). Comparison of ``GPT'' potentials (Ref.~\onlinecite{moriarty-GPT}) 
with empirical ``EOPP'' potentials based on ab-initio database, 
both fitted to form (\ref{eq:oscil6}).
The Cu--Sc potential is not available from Ref.~\onlinecite{widom-GPT};
the figure shows a fit to ab--initio data with the other five pairs
constrained to GPT form. The ``GPT''/fitted curves are shifted by $\pm0.1$ 
eV/atom for clarity. 
}
(b). Scatter plot of the forces 
for the EOPP fit to [Eq.~(\ref{eq:oscil6})], from the
fitted Al-Cu-Sc potentials shown in (a), bottom. 
The pair potential forces ${\bf F}_j$   are vertical
axis, ab-initio data horizontal.
}
\label{fig:pots}
\end{figure}

We define each structure's energy as a difference
relative to a coexisting mixture (with the same 
total composition) of reference phases,
chosen to bracket all database compositions.
\SAVE{Depending what tie-plane you
choose, you obtain different potentials.  
MM says that, by choosing the tie-plane to bracket the
database (usually, pure elements), you reproduce the energy part
of the database better.
(What would implement a change of chemical potentials?  Does a 
change in tie-plane do that?)
MW: ``Rather than adding constants, its better to think about adding a
localized term that alters the near-neighbor energy without
significantly altering the near-neighbor force. For pure elements
this amounts to a chemical potential shift but for mixed
potentials it governs the enthalpy of mixing.
Of course the potentials [as in (\ref{eq:oscil6})]
need to [asymptote] to zero to assure convergence of the energy.''}
Every structure is used for both forces (from MD at high $T$) and
energy differences~\cite{FN-need-energies}
(high-$T$ MD, as well as relaxed at $T=0$).
For the high-$T$ portion, we took one snapshot of each structure
at the end of a short ab-initio MD run.
Usually $\gtrsim 10^3$  force components enter the fit, 
along with $\sim 50$ energy differences.
Typical forces are $\sim 3$ eV/\AA~ and 
typical energy differences are $\sim 0.3$ eV/atom;
the fit residuals are $\sim 5$\% and $\sim 1$\% respectively
[see e.g. Fig.~\ref{fig:pots}(a)].

Our least-squares fit minimizes (by the Levenberg-Marquardt algorithm)
     $\chi^2 \equiv \sum \Delta E_i^2/ {\sigma_E}^2 + 
     \sum |\Delta {\bf F}_j|^2/{\sigma_F}^2$, 
where $\{ \Delta E_i \}$ and $\{ \Delta {\bf F}_j \}$ are the
energy and force residuals; we found a weighting ratio
$\sigma_E/\sigma_F \sim 10^{-3}$\AA~ was optimal so
that neither energies nor forces dominate the fit.

There is some risk of converging to a false minimum
(or not converging at all, from an unreasonable initial guess).
Thus, it is important to repeat the fit from several starting guesses.  
For this we used, e.g., potentials first fitted to pure elements
or binary systems, and also used a library of parameter sets
previously fitted for some different alloy system.
The fitted parameters in (\ref{eq:oscil6}) for each of our 
examples are available as supplementary information~\cite{suppl};
similar potentials were plotted in Refs.~\onlinecite{CMA-AlMg} (for Al-Mg) 
and \onlinecite{boissieu-ScZn} (for Sc-Zn).

For our specific case of Al--Cu--Sc, our database had 84
relaxed energies at $T=0$K from (besides the pure elements) the binaries
Al$_3$Sc.cP4, Cu$_2$Sc.tI6, Al$_2$Sc.cF24, and Al$_2$Cu.cF12 and
ternaries  AlCuSc.hP12,  Mg$_2$Cu$_6$Ga$_5$.cP39 (with Mg$\to$Sc, Ga$\to$Al),
and AlCrCu$_2$.cF16 (with Cr$\to$Sc); structures are identified by
Pearson labels.  The database also had 7428 force points 
at $T>0$K,  taken from all those structures, and additionally 
from Al$_2$Sc.hP12, and Sc(Al$_{1-x}$Cu$_{x})_6$.cI168.
The EOPP potentials give very good
agreement with this database,
as shown in Figure~\ref{fig:pots}(a):
the r.m.s. deviation of forces was 0.11 eV/\AA~ 
and that of energy differences was 21.7 meV/atom, with
relative weights set to maximize accuracy of the force
data.~\cite{FN-weighting} Values of the parameters are listed
in Table~\ref{tab:o6-alcusc}, Appendix~\ref{append:eopp}.
\SAVE{(The energies are,  in fact, formation enthalpies,
since we expressed the energy datapoints relative to 
the pure elemental energies tie--plane).
Note that this fit is rather biased towards the forces;
by using the same data but increasing the relative weight for the energies
relative to the forces,
we could decrease r.m.s.(energy) to $\sim$10 meV/atom, at the price that  r.m.s.(forces) 
increases to 0.135eV/\AA.  Table I in an earlier draft quoted r.m.s.(forces) was 0.15 eV/\AA~ and
r.m.s.(energies) was 3.4 meV/atom. This presumably refers to an older version of
the database, and/or a fit weighted even more strongly towards energies.}

Our empirical potentials can be compared with 
Moriarty's ``GPT'' potentials ~\cite{moriarty-GPT},
derived from a systematic expansion, which are known 
for all but one of the six Al--Cu--Sc pair potentials
[see Fig.~\ref{fig:pots}(b)].
First, when the GPT potentials are fitted to the EOPP form 
(Eq.~(\ref{eq:oscil6}), the curves virtually lie on top of each other:
thus, the EOPP form {\it can} represent all features of the GPT potentials. 
Second, the GPT and EOPP potentials
show a strong similarity; the main mismatches are (i) EOPP has no
first-neighbor well for Al--Cu and Cu--Cu (ii) the oscillation
wavelength for Sc-Sc differs (iii) the overall energy scale of 
the GPT pair potential 
is too small by $\sim$50\%.~\cite{FN-Al-TM}.
In practice empirically fitted potentials account for some
of the many-body contributions by modifying their pair terms, 
and hence work better than truncating a systematic expansion 
(e.g. GPT~\cite{moriarty-GPT}) 
after the pair terms.

\secc{Constrained-cell melt quenching}
\label{sec:constrained}
We now turn to the second method which, together with fitted
potentials, has enabled the present structure study and others.
In many alloy systems, with no structural information known except
the composition {\it and the unit cell}, structures with  $>100$
atoms per cell may be predicted from careful ``melt quenching''
(MQ) simulations.  The relation to the EOPP notion is
that the larger systems -- particularly when supercells are used 
-- are too large for direct ab-initio calculations,
so empirical potentials are crucial for melt-quenching.
This method has been applied with GPT potentials to improve
known structures of the decagonal 
Al--Co--Ni quasicrystals~\cite{MQ-AlCoNi}.

\SAVE{The ``cell-constrained'' method is attractive when
single-phase crystals cannot be prepared, so that
even powder diffraction is infeasible.}

In most cases, annealing requires
``tempering Monte Carlo (MC)''~\cite{TEMPERING} 
wherein $\sim 10$ samples are annealed simultaneously 
at equally spaced temperatures spanning across the
melting temperature.
\SAVE{(More specifically, spanning from about 1/2 the melting temperature to above it,
and spaced by $\Delta T \approx$ 50K.)}
Each tempering cycle includes a short
MD run ($\sim$ 1 ps with 1 fs steps)
followed by lattice-gas MC annealing 
($\sim 200$ attempts per atom) 
in which the chemical identities of two randomly
selected atoms may be swapped.
Then, pairs of samples may be
swapped using a Metropolis-like criterion.
(For the Al--Cu--Sc structures we studied,
our temperatures spanned 600--1700 K with spacing $\Delta T = $100K,
and the interaction cutoff was set to r$_{\rm cut}=9$\AA.)
\OMITX{(For the 39-atom $cP39$ structure, they spanned 600K--1650K with
$\Delta T=$ 150K.)}
The resulting structure may be tested subsequently by 
ab-initio calculations of the total energy, for which we
used VASP~\cite{vasp}.

A key diagnostic in a tempering simulation is 
the (time-dependent) energies $E_m(t)$ of all samples,
as specific heat is approximately 
$\Delta E/\Delta  T$, where  $\Delta E=E_{m+1}-E_m$.
For our structures  $\Delta E$ 
peaked around $T=$1400K, indicating the melting point.
In our cases with 52 or 84 atoms per primitive cell,
{\it several} independently initialized runs 
yielded identical low-$T$ structures,
which took 100--200 cycles.
\OMITX{($oC104$ and $cI168$, see below;
only 20--30 cycles were needed for $cP39$.)}

\secc{Target structures}
We now apply empirical pair potentials and cell-constrained 
melt-quenching to realistic structure prediction in the Al--Cu--Sc system,
for two newly discovered phases~\cite{ishimasa-oC104,ishimasa-cI168}
in which lattice parameters were known from electron microscopy but
no single-grain structures were available for structure determination.
\OMITX{We also found a third, hypothetical phase that is expected to be 
(nearly) stable.}
We label each phase by its Pearson symbol.
\SAVE{Each phase is an ``approximant'' of a quasicrystal phase.
Note the phases of Al-Cu-Sc alloys have structures similar to the
ZnSc alloys with Zn $\to$ (Al,Cu).}
These belong to a family of phases in which atom 
size is the salient attribute:  Ca, Mg, Sc, and rare earths 
are ``large'' atoms, constituting $\sim 1/6$ by number;
whereas Al, Zn, Cu, and Cd are the majority ``small'' atoms.

Our first target is ``$cI168$'' with 84 atoms per primitive cell,
hypothesized to be isostructural with the ScZn$_6$ phase. 
No single-crystal data are available; a 
preliminary Rietveld refinement of powder data technique~\cite{ishimasa-cI168}
confirmed the ScZn$_6$ type structure, with refined lattice parameter
$a$=13.52\AA.  
The chemical ordering and occupancies of Al/Cu could not be determined reliably
prior to our modeling~\cite{ishimasa-cI168}.

The ScZn$_6$ phase is an ``approximant'' of 
the recently discovered thermodynamically stable,
icosahedral quasicrystal $i$(ScZn),
meaning its unit cell contents are identical 
to a fragment of the quasicrystal structure.
$i$(ScZn) has the same atomic structure as $i$(CaCd)~\cite{tsai-CaCd},
which (along with similar rare earth-Cd quasicrystals) are
the only known stable binary quasicrystals. Each site is
specific to either a small atom (Cd or Zn) or a large 
atom (Sc, Ca, or rare earth).

\begin{figure}
\vskip 1.0cm   
\includegraphics[width=3.2in]{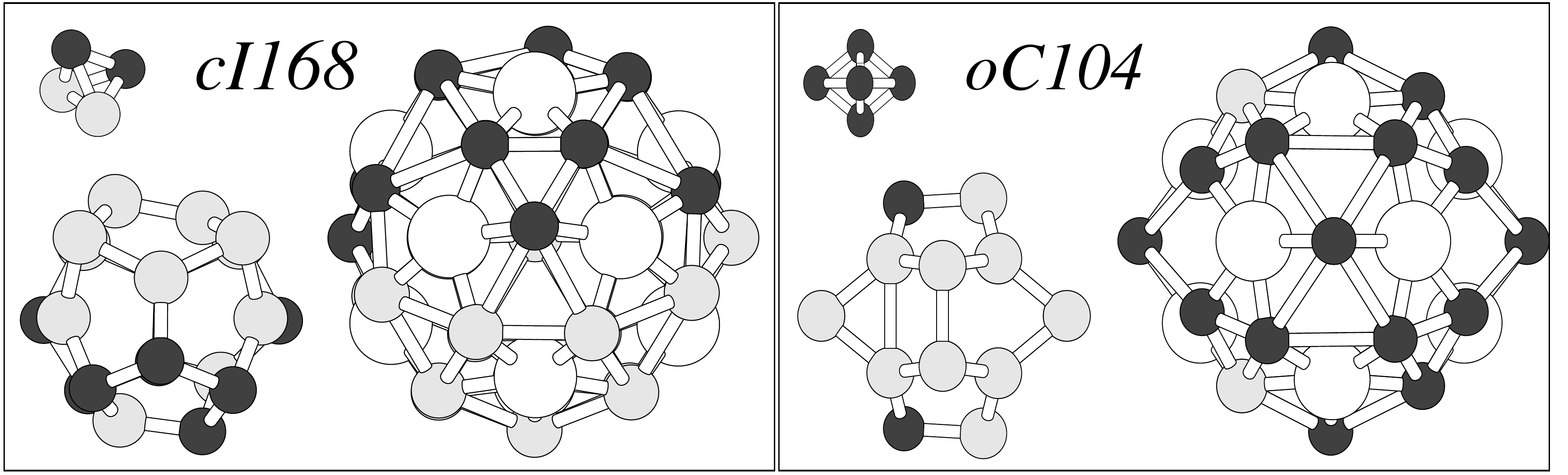}
\caption{
(a)  Tsai cluster in $cI168$ structure (left) and 
(b) ``pseudo-Tsai'' cluster (right) found in the $oC104$
  structure. View along 2--fold direction.
In each panel, the successive shells are (1) tetrahedron Al$_2$Cu$_2$ or
  octahedron Cu$_6$ (2)  dodecahedron (Al,Cu)$_{20}$, and (3)
  icosidodecahedron (Al,Cu)$_{30}$ plus icosahedron Sc$_{12}$.
Large white circles represent Sc, dark smaller circles Cu, 
and light smaller circles Al.
\SAVE{
[The pseudo-Tsai cluster in hypothetical Sc$_2$Cu$_{6+x}$Al$_{5-x}$.$cP39$ 
is identical to (b)
except for the replacement Al$\to$Cu on two inner and two
outer sites (shown by arrows).]
}
}
\label{fig:clusters}
\end{figure}

Both $cI168$-ScZn$_6$ and the related quasicrystals are understood
to be packings of a three-shell,
icosahedrally symmetric cluster called the 
``Tsai cluster''. Its outermost shell has an icosahedron
of 12 large atoms plus 30 small atoms (forming an icosidodecahedron)
on the midpoints of the large-large bonds [See Fig.~\ref{fig:clusters}(a)],
and inside that is a dodecahedron of 20 small atoms. 
At the cluster center is a {\it tetrahedron} of small atoms;
as this breaks the icosahedral symmetry, it has many
degenerate orientations, which leads to interesting 
slow dynamics of reorientations~\cite{ScZn-tet} and
to structural ordering transitions at low $T$ in 
the related crystals.

Our second target is ``$oC104$'', 
of composition Al$_{38.8}$Cu$_{45.7}$Sc$_{15.5}$,
\SAVE{The structure did not have any obvious prototype.}
Initially ~\cite{ishimasa-oC104} the lattice parameters 
$a$=8.32\AA, $b$=8.36\AA, $c$=21.99\AA~and 
C-centered orthorhombic Bravais lattice were identified from
powder data but that was insufficient to refine the structure.

\secc{Results: phase stability and pseudo-Tsai clusters}
\label{sec:results}
Constrained-cell melt-quenching  using the same database-fitted EOPP potentials 
converged to low--temperature structures in both cases, the energies of which 
were subsequently evaluated by VASP. 
In the $cI168$ case this was the known structure.
\OMITX{
In the $cP39$ case with composition Sc$_2$Cu$_6$Al$_5$,
melt-quenching recovered the Mg$_2$Zn$_{11}$ structure with
(as had been guessed) 
exactly the chemical ordering of Mg$_2$Cu$_6$Ga$_5$ as found in
Ref.~\onlinecite{lin-CuGaMg},
with Mg$\rightarrow$Sc and Ga$\rightarrow$Al);
the $cP39$ structure is stable only after substituting Ga$\to$Cu on
one or two Ga(2) sites, i.e. for $x=1/3$ or $2/3$.
}
In the $oC104$ case, a previously unknown structure was obtained~\cite{suppl},
computed unstable by only 5 meV/atom with respect to $cI168$; its validity
was confirmed by a subsequent refinement of the powder data~\cite{ishimasa-oC104} 
with this as the starting point.
\SAVE{
(Clipped footnote about oC104).
We tried modifications with doubled cell, Al$\to$Cu substitution,
or lower atomic density, but it turned out that the optimum variant 
of the $oC104$  structure has exactly the same composition and 
number density as $cP39$ (which in turn was the same as Mg$_2$Zn$_{11}$).
Even in the large-cell $cI168$ case, melt-quenching converged 
to a $T=0$ stable structure, with composition
Al$_{34}$Cu$_{38}$Sc$_{12}$.
}
This predicted structure had space group $Amm2$,
in which there were 11 Cu, 7 Al and 4 Sc Wyckoff sites, with
no chemical disorder.  

The fundamental motif in both $cI168$ and $oC104$
is an icosahedral cluster whose
innermost shell has less symmetry.  In $cI168$ this is a Tsai cluster
[Fig.~\ref{fig:clusters} (a)] in which 
each of the three Al/Cu shells has composition Al$_{0.5}$Cu$_{0.5}$.  
Strikingly, within each shell,
the Al and Cu are segregated into hemispheres centered on
fivefold axis of the icosahedron~\cite{FN-pseudo5};
furthermore, the Al parts of each shell overlay the Cu
parts of the preceding one and vice versa 
[see Figure~\ref{fig:clusters}(a)] .
\SAVE{The innermost shell is an Al$_2$Cu$_2$ tetrahedron; 
its Cu$_2$ bond aligns with
the Al hemisphere of the (second shell) dodecahedron,
followed by Cu--hemisphere of the (third shell) icosidodecahedron).}
Due to this symmetry breaking, one expects the 
low--temperature phase to have at least several equivalent domains,
corresponding to different orientations of the pseudo-Tsai
clusters.

\SAVE{The calculated Tsai clusters in $cI168$ agree well
with the Rietveld refinement~\cite{ishimasa-cI168}:
atom radii are within $\sim 2$\%, and the fraction
of Cu (on Al/Cu sites) is also within a few percent,
except too much Cu is predicted on the inner tetrahedron.}

\MEMO{QQ-1. Do we need to add anything here? See these remarks.
MM remarks: I find it intriguing that the experimentally
fitted Al/Cu occupancies for these shells are almost exactly 0.5Al:0.5Cu!
CLH remarks: Why is it aligned along a fivefold axis?
MM remarks: we don't know interactions of these cluster orientations.
MM: No!}
In the $oC104$ case we find a variant motif~\cite{euchner}
that we call the ``pseudo-Tsai'' cluster, in which 
the innermost shell is a Cu$_6$ octahedron [Fig.\ref{fig:clusters}(b)].
This cluster was first described in Mg$_2$Zn$_{11}$.$cP39$~\cite{euchner},
where the outer shell is 
a Zn$_8$Mg$_{12}$ dodecahedron plus a Zn$_{12}$  icosahedron~\cite{cp39}.

\OMITX{Actually $cP39$ is a packing of {\it two} kinds of icosahedral clusters
(sharing their outer shells): a pseudo-Tsai cluster sits at
each cell center, while at each cell corner sits a 
``Pauling triacontahedron''.
The latter consists of the inner part of a 
Bergman cluster~\cite{bergman}): in contains 
a center site (fully occupied Al in the AlCuSc version of $cP39$),
an inner Cu$_{12}$  icosahedron, and a 32-atom outer shell.}

The experimental (powder data) refinement~\cite{ishimasa-oC104}
of $oC104$ is an {\em average} structure, with
higher symmetry (space group $Cmmm$) than our model, accomodating
3 Sc Wyckoff sites, and 13 mixed Al/Cu sites (one of the latter
being half-occupied). Each experimental site derives from one or
two sites in our model, displaced on average by $\sim$0.17\AA~ (Al),
$\sim$0.12\AA (Cu) and 0.08\AA (Sc). Crystallographic data of the
low--temperature model structure, are compared with the 
experimental structure in Appendix~\ref{append:oc104}.
\MEMO{QQ-2
``(except the two sites adjacent to the half--occupied site, 
have 0.3-0.4\AA~ displacements)'': I would change and quote
the r.m.s. mean displacement averaged over all sites.}
\MEMO{MM: Well, about half of the sites
have a SINGLE model site corresponding to it, the other half has a PAIR.
From the model, the possible average occupancies are: pure Al, pure Cu, or
0.5Al+0.5Cu (two sites). The expe chemistries are always correct in terms
of ``majority'' when compared to the pure Al or pure Cu model sites. And they
are not far in the two cases when the averaged model site is 0.5Al+0.5Cu...
Could you please try to condense this information? 
REPLY: The brief way, in my opinion, is to quote averages.
Can you supply them?
MMrep : provided averages, see text.}
The mean deviation of the refined Al/Cu content from the averaged
model content is 16\%.

\SAVE{In percentages, this is Al$_{40.5}$Cu$_{45.2}$Sc$_{14.3}$.
Ishimasa's 2009 abstract for the Physical Soc. of Japan gives
refined partial occupancies adding up to 
Al$_{32.8}$Cu$_{39.1}$Sc$_{12}$.}

The electronic density of states for $cI168$
has a deep, narrow pseudogap at the Fermi energy,
which tends to stabilize a unique composition. 
In contrast, $oC104$ has a shallower and wider pseudogap,
suggesting a range of degenerate compositions, 
so that substitutional entropy
might stabilize this phase at higher temperatures.

The pseudo-Tsai clusters in $oC104$ (and also 
the less complex $cP39$) adjoin by sharing atoms
such that their centers are closer by a factor
$\sim 1.618$ (golden ratio) than Tsai clusters would be.
Could pseudo-Tsai clusters be the basis of
the newly reported i(AlCuSc) quasicrystal~\cite{ishimasa-iAlCuSc}?

\secc{Conclusion.}
We have shown that empirical potentials with the simple oscillating form 
(\ref{eq:oscil6}), fitted from ab-initio data and
combined with a ``cell constrained'' brute-force quenching,
allows detailed predictions of fairly complex low-temperature 
optimal structures in Al--Cu--Sc alloys, 
based on the very limited input of known lattice parameters
and  composition.
Finding the correct structure depends 
sensitively on having a quantitatively realistic potential,
which is achievable only if that potential is constructed
or fitted from ab-initio calculations.
One would expect that the oscillating analytical form (\ref{eq:oscil6}) 
is natural only for simple metals (e.g. Al or Mg, for which it does very well).
But in fact, the EOP potentials sometimes work quite well 
even when angular or many-body interactions are important, 
e.g. the transition metal neighbors in Al--Cu--Sc.  
But -- not surprisingly -- they do poorly for elemental Zn or Ga.
\MEMO{QQ-3 Actually, why is it not surprising for Ga?  
(I was wondering if the remark about Zn and Ga belongs
after the next sentence, which talks of covalency.) 
MM: I think it's that both Zn and Ga have d-bands very near Fermi energy. 
CLH: So is that bad because d-bands are covalent?  Also,
if that happens why aren't they magnetic?
MMrep: I'm not certain... forgot to ask Marian today.
}
\SAVE{Pair potentials never reproduced the ab-initio phonon vibrational DOS of Zn.
In pure Zn, pair potentials fail even when using better functional 
forms than  (\ref{eq:oscil6}).
The EOPP approach just fails for pure Ga.}
Of course, {\it any} pair potential fails when the electron density has 
large variations in space (as at vacancies, edge dislocations, or surfaces), 
or in the (many) cases where bond directionality (due to covalent bonding)
is  prominent.

We believe the EOPP potentials are quite broadly applicable
to mimic the atomic interactions of many metallic
systems with sufficient accuracy to stand in for
ab-initio energies when those would be computationally prohibitive.
Although the EOP potentials were not formally presented before
this paper, they have already been applied to a variety of 
intermetallics:
(1) the site contents in complex structures, e.g. 
Al--Mg~\cite{CMA-AlMg} or Al--Zn--Mg~\cite{boissieu-AlZnMg};
(2) solving the complete structures of complex 
Mg-rich Mg--Pd phases (with $>400$ atoms/cell)
in conjunction with diffraction, when the
latter alone is insufficient~\cite{MM-kreiner};
(3) phonon spectra, e.g.  in 
Zn--Sc~\cite{boissieu-ScZn} 
and Mg--Zn~\cite{euchner} alloys, and even in 
liquid Bi-Li alloys~\cite{Bi-Li};
(4) the dynamics of the Tsai cluster tetrahedra 
in the $cI168$ structure ScZn$_6$~\cite{ScZn-tet},
as well as the arrangements of the asymmetric 
inner Al$_{10}$ shell in the pseudo-Mackay icosahedral clusters in
quasicrystal-related Al--Ir, Al--Pd--Mn, 
and Al--Cu--Fe phases~\cite{AlIr-AlPdMn}.

\acknowledgments
We thank R. G. Hennig and T. Ishimasa for discussions, and
M. Widom for collaboration at an earlier stage.
This work was supported by DOE Grant DE-FG02-89ER-45405
(MM, CLH), and Slovak funding VEGA 2/0111/11 and APVV-0647-10 (MM).

\appendix

\secc{EOPP parameters for Al--Cu--Sc system.}
\label{append:eopp}

The EOPP potential parameters fitting Eq.\ref{eq:oscil6} to 
DFT/VASP data are shown in Table~\ref{tab:o6-alcusc}.

\begin{table}[p]
\begin{tabular}{c|cccccc}
\hline
 & $C_1$ & $\eta_1$ & $C_2$ & $\eta_2$ & $k_*$ & $\Phi_*$ \\ 
\hline
Al--Al &  372.35 & 7.310 & -1.508 & 3.600 & 3.552 & 2.995  \\
Al--Sc &  459191.27 & 15.730 & -17.912 & 4.656 & 3.363 & 1.693  \\
Al--Cu &  249.79 & 7.563 & -1.563 & 3.025 & 2.937 & 5.812  \\
Sc--Sc &  18493.28 & 11.264 & 2.379 & 2.559 & 2.308 & 0.947  \\
Sc--Cu &  194.11 & 7.361 & -10.462 & 4.255 & 2.721 & 4.548  \\
Cu--Cu &  1036.62 & 9.503 & -0.875 & 2.818 & 3.095 & 6.052  \\
\end{tabular}
\caption{\label{tab:o6-alcusc} Fitted parameters for Al--Cu--Sc EOPP potentials.}
\end{table}

%
%

\secc{Crystallographic data of the $oC104$ structure}
\label{append:oc104}

Table~\ref{tab:struc} lists Wyckoff orbits of the low--temperature
structure, resulting from the melt--quenching procedure described in 
Sec.~\ref{sec:constrained}, using EOPP Al--Cu--Sc potentials
(Tab.~\ref{tab:o6-alcusc}). The structure was subsequently optimized
using DFT code VASP.  The experimental (relaxed) cell parameters were
$a$=8.3370 (8.3247) \AA, $b$=22.0150 (21.868) \AA~and $c$=8.3370
(8.3247) \AA, space group $Amm2$ (no. 38).  Note that $a$ and $c$ axis
are swapped with respect to the setting in
Ref.~\onlinecite{ishimasa-oC104}. Column $\mu$ is site multiplicity,
column $\Delta$R shows displacement of the final relaxed atomic
positions from the refined diffraction--data sites, and the last
column (x$_{Al}$) is occupancy by Al atom from
Ref.~\onlinecite{ishimasa-oC104}. The experimentally determined
structure has higher symmetry (space group $Cmmm$), and shows
averaging over disorder manifested at M2 site, forming 0.8\AA--distant pairs of
half--occupied positions.
Occupancy of the M2 site couples with Al/Cu symmetry breaking of
its adjacent M8a/b and M9a/b sites -- all other sites marked a/b in the table
(Sc3, M3, M10 and M13) have unique chemical assignement. Not surprisingly,
largest displacement $\Delta$R from the $Cmmm$ diffraction--data model occurs for
Al--occupied variants of the M8 and M9 sites (Cu is stronger X--ray scatterer).
The average $Cmmm$ model can be therefore interpreted as thermal average over
(locally) ordered patterns, in which flip of the M2 atom by
$\sim$0.8~\AA ~over pseudo--mirror plane perpendicular to $c$ axis at $z$=0
is correlated with Al$\equiv$Cu swap of adjacent M8a/M8b and M9a/M9b sites.

\begin{table}
\small
\begin{tabular}{c|cc|ccc|cc}
\hline
site  & chem. & $\mu$& X  & Y  & Z   & $\Delta$R  & x$_{Al}$\\ 
Sc1      & Sc & 8  &    0.6942 & 0.8140 &    0      & 0.10 & --  \\ %
Sc2      & Sc & 4  &    0      & 0.3844 &    0.0008 & 0.03 & --  \\
Sc3a     & Sc & 2  &    0.5    & 0      &    0.3051 & 0.04 & --  \\
Sc3b     & Sc & 2  &    0.5    & 0      &    0.3086 & 0.07 & --  \\
M1       & Cu & 4  &    0.2564 & 0      &   -0.0064 & 0.07 & 0.05 \\
M2$^\dag$& Cu & 4  &    0      & 0.2574 &    0.0414 & 0.03 & 0.96 \\  
M3a      & Cu & 2  &    0      & 0      &    0.2452 & 0.09 & 0.11 \\
M3b      & Cu & 2  &    0      & 0.5    &    0.2322 & 0.09 & 0.11 \\
M4       & Cu & 8  &    0.3465 & 0.4066 &   -0.0046 & 0.04 & 0.12 \\
M5       & Al & 4  &    0.5    & 0.0687 &   -0.0014 & 0.04 & 0.88 \\
M6       & Cu & 4  &    0      & 0.0661 &   -0.0152 & 0.17 & 0.22 \\
M7       & Cu & 4  &    0.5    & 0.2490 &    0.2537 & 0.04 & 0.68 \\
M8a      & Cu & 8  &    0.2408 & 0.3081 &    0.1583 & 0.05 & 0.28 \\
M8b      & Al & 8  &    0.2144 & 0.2963 &   -0.1652 & 0.30 & 0.28 \\
M9a      & Cu & 4  &    0      & 0.3446 &   -0.3383 & 0.01 & 0.15 \\
M9b      & Al & 4  &    0      & 0.3240 &    0.3334 & 0.46 & 0.15 \\
M10a     & Cu & 4  &    0.5    & 0.3689 &    0.2544 & 0.03 &  0  \\
M10b     & Cu & 4  &    0.5    & 0.3684 &   -0.2560 & 0.01 &  0  \\
M11      & Al & 4  &    0.1821 & 0.5    &   -0.0072 & 0.06 & 0.78 \\
M12      & Al & 4  &    0.5    & 0.3050 &    0.0034 & 0.03 & 0.69 \\
M13a     & Al & 8  &    0.2201 & 0.4187 &    0.2728 & 0.08 & 0.93 \\
M13b     & Al & 8  &    0.2194 & 0.4199 &   -0.2820 & 0.07 & 0.93 \\
\hline
\end{tabular}\\
$^\dag$ this site has 0.5 occupancy in the refinement Ref.~\onlinecite{ishimasa-oC104}.\\
\caption{\label{tab:struc} List of Wyckoff sites for $oC104$ structure
  minimizing EOPP energy, starting from random initial state, and
  using fixed experimentally determined lattice parameters as the only
  input. Last two columns compare the refined structure with the Rietveld--refined
structure of Ref.~\onlinecite{ishimasa-oC104}, see text.}
\end{table}


\begin{thebibliography}{99}
\bibitem{FN-empirical}
``Empirical potentials'' are explicit, classical functions 
of coordinates, not requiring an eigenvalue calculation.
\bibitem{FN-weighting}
An alternative weighting favoring accuracy in the energies 
would decrease the r.m.s. deviation for the energy datapoints 
to 7.9 meV/atom, while increasing the r.m.s. deviation
of the force data to 0.15 eV/\AA.
\bibitem{FN-need-energies}
Fits from forces alone were insufficiently constrained;
including energy differences, even with low weight,
improved the fit.
\bibitem{FN-pseudo5}
Properly ``pseudo-fivefold'', as this is in a cubic
crystal.
\bibitem{cp39} In fact we find ternary variant of
Mg$_2$Zn$_{11}$.cP39, the Mg$_2$Cu$_6$Al$_5$.cP39 structure stable,
coexisting with the $cP168$ structure, at T=0K.
\end{thebibliography}
\end{document}